\documentclass[a4paper]{article}

\usepackage{icrc2013,xspace}

\title{Impact of E-ELT laser light on Cherenkov Telescope Array cameras}

\shorttitle{Impact of E-ELT lasers on the CTA}

\authors{
M. Gaug$^{1,2}$,
M. Doro$^{1,2,3}$,
for the CTA Consortium.
}

\afiliations{
$^1$ F{\'i}sica de les Radiacions, Departament de F{\'i}sica, Universitat Aut{\`o}noma de Barcelona, 08193 Bellaterra, Spain. \\
$^2$ CERES, Universitat Aut{\`o}noma de Barcelona-IEEC, 08193 Bellaterra, Spain. \\
$^3$ University and INFN Padova, Via Marzolo 8, 35131 Padova, Italy. \\
}

\email{markus.gaug@uab.cat}

\abstract{As one of the options, the Cherenkov Telescope Array (CTA) Consortium is
considering the possibility to install its Southern array in Chile,
in the Atacama Desert. The envisaged site is situated about 5 km from
the future European Extremely Large Telescope (E-ELT), which will operate
8 parallel DC lasers emitting at 589.2 nm, to create an artificial 6.8
magnitude star at an altitude of 90~km. The guide stars are used for the
adaptive optics of the telescope. Although having the artificial stars in
the field-of-view of a CTA telescope would happen rather seldom, and can
be avoided by coordinated scheduling, the laser beams may cross the
field-of-view of a telescope more frequently and leave spurious light
tracks, hence complicating the analysis of the shower images.  
We derive an approximate formula to estimate the expected number of photons 
from molecular and aerosol scattering of the laser light beam into the field-of-view of a
camera pixel. We then present several specific cases of laser influence on
the CTA camera pixels, based on the selected direction of the laser beam,
using the expected quantum efficiency of the camera photomultipliers at
the given wavelength.}


\begin{document}
\maketitle

\section{Introduction}
Chile is now being considered seriously as a possible host country for
the Southern array of the CTA ~\cite{ctaconcept}. If accepted, the array would be
installed in the {\it Paranal - Cerro  Armazones} plateau, close to
the location of the future \textit{European Extremely Large Telescope} (E-ELT), 
which belongs to the {\it European Southern Observatory} (ESO).   
The most promising candidate site lies at a
distance of about 5~km from the \mbox{E-ELT}, taking advantage of the
infrastructure built for that installation. The E-ELT
will operate 8 powerful DC lasers to create artificial guide-stars for
the adaptive optics of its primary mirror.  These lasers will operate
at elevations of as low as 20$^\circ$ and could therefore cross the
field-of-view of the CTA telescopes nearby. The reflected laser
light could then leave spurious light tracks in the cameras, affecting
the analysis of the shower images or possibly triggering fake events.
  
We present here formulae to perform quick calculations for the amount
of expected light from the laser beams in any of the CTA cameras, for
a given distance to the E-ELT lasers and laser pointing
angles. Inserting concrete values for typical case scenarios, and one
absolutely worst case, will yield absolute numbers which can then be
compared with other sources of background light, such as faint stars
and/or the night-sky background.  

\section{The lasers}
The E-ELT will operate 8 DC, extremely well collimated, lasers at a wavelength of 589.2~nm, each
with a power of 20~W. The lasers will excite a layer of sodium atoms
in the mesosphere (reaching about 90~km altitude a.s.l. for vertical
shots) which then re-emit the laser-light and appear to ``glow''.  
These so-called {\it sodium laser guide stars} most likely use 
circularly polarized laser light to achieve maximum impact~\cite{Holzloehner}. 
The following table provides the relevant data: 
\begin{center}
\vspace{-0.4cm}
\begin{tabular}{lll}
Parameter  &  Value & Comments \\
\hline
Number of lasers & 8    & Fired in parallel   \rule{0mm}{3mm}  \\
DC power         & 20 W & $5.9\cdot 10^{19}$~ph/s/laser \rule{0mm}{3mm} \\
Max. distance    & 42 m &    \rule{0mm}{3mm}  \\
betw. lasers     &      &                                             \\
Distance on ground  & 5~km &  depends on exact   \rule{0mm}{3mm} \\
to the CTA       &      &  location of the CTA \\
Beam width       & 0.5~m &                \rule{0mm}{3mm} \\
Beam opening     & 0.01~mrad &                \rule{0mm}{3mm} \\
 angle           &      &                      \\
Operation elevation  & 20--90$^\circ$ & mostly above 45$^\circ$   \rule{0mm}{3mm} \\
Altitude        &  3~km a.s.l. &   \\
\hline
\end{tabular}
\end{center}

\section{Scattering of the laser light}
First, we consider Rayleigh scattering of light in dry air. Light
of wavelength $\lambda$ and polarization angle $\phi$ is 
scattered by air molecules at a scattering angle $\theta$ with respect
to the direction from which the photon impinges, with the
following cross section~\cite{Penndorf1957}: 
\begin{eqnarray}
\frac{d\sigma(\phi,\theta,\lambda)}{d\Omega} &=& \frac{9\pi^2 \cdot (n^2-1)^2 }{\lambda^4 \cdot N_s^2 \cdot (n+2)^2} \cdot \big(\frac{6+3\rho}{6-7\rho}\big) \cdot \nonumber\\
     && \cdot \big( \cos^2(\phi)\cos^2(\theta) + \sin^2(\phi) \big)~. 
\end{eqnarray}
In the above formula, $N_s$ is the molecular concentration, $n$~the
refractive index of air and $\rho$ the de-polarization ratio. Note that
since $(n^2-1)/(n+2)$ is proportional to $N_s$, the resulting
expression depends only on the particle mixture, and is independent of
particle density as well as temperature and pressure~\cite{Bodhaine}. Hence we can
pick one reference condition for ($T,P$), which is typically made for
the standard reference case $T_s=288.15$~K and $P_s=1013.25$~mbar. \\
$N_s$ is then
$2.547\cdot 10^{25}$~m$^{-3}$. The combination $(n^2-1)/(n+2)$ yields
 $1.84\times 10^{-4}$ at $\lambda = 589$~nm~\cite{Peck1972}.  
The \textit{King factor} $(6+3\rho)/(6-7\rho)$ describes the effect of
molecular anisotropy and amounts to about 1.05~\cite{tomasi}. 
Multiplying with the number density of molecules at a given
height~$h$, we obtain the \emph{volume scatter coefficient}
$\beta(\lambda,\theta,\phi,h)$: 
\begin{eqnarray}
\beta_\mathrm{mol}(589~\mathrm{nm},\theta,\phi,h) &\approx& 10^{-6}\cdot \nonumber\\
&& \cdot \big( \cos^2(\phi)\cos^2(\theta) + \sin^2(\phi) \big)  \cdot \nonumber\\
&&\cdot \frac{N(h)}{N_s}~\mathrm{m^{-1}\,sr^{-1}}~. 
\end{eqnarray}
Assuming \textit{un-polarized light}, or a \textit{circularly polarized
  light} beam seen over a field-of-view much larger than the
wavenumber of a 589~nm light wave, the equation reduces to: 
\begin{equation}
\beta_\mathrm{mol}(589~\mathrm{nm},\theta,h) \approx 10^{-6} \cdot \frac{ \cos^2(\theta) + 1 }{2}
\cdot \frac{N(h)}{N_s}~\mathrm{m^{-1}\,sr^{-1}}~. 
\end{equation}
We then consider a US standard troposphere~\cite{USStandard} with:
\begin{equation}
\frac{N(h)}{N_s} \approx (1 - 2.3\cdot 10^{-5} \cdot h)^{4.256}~.
%\frac{N(h)}{N_s} = (1 - \frac{L\cdot h}{T_s})^{\frac{g\cdot M}{R \cdot L}-1} \approx (1 - 2.3\cdot 10^{-5} \cdot h)^{4.256}~,
\end{equation}
%
%where $L=0.0065$~K/m is the temperature lapse rate of the troposphere,
%$T_s=288.15$~K the standard temperature at Sea level,
%$g=9.8~\mathrm{m}\cdot\mathrm{s}^{-2}$ the gravitational constant,  
%$M=0.029$~kg/mol the molecular weight of dry air and
%$R=8.31~\mathrm{J}\cdot\mathrm{mol}^{-1}\cdot\mathrm{K}^{-1}$ the
%ideal gas constant. \\

%\subsection{Mie Scattering}
\noindent
Aerosols scatter light more efficiently than molecules, due to their
larger sizes. In order to reliably estimate the effect of aerosols
on the scattering of the laser light, their
size distribution and height dependency need to be known. We have not
found any aerosol model for Paranal; however, the optical extinction has been described in great detail by~\cite{patat}. 
Aerosol extinction at 589~nm is found to be $k_\mathrm{aer}(589~\mathrm{nm})=0.027\pm0.004(\mathrm{stat.})\pm0.004(\mathrm{syst.})$, 
where the systematic uncertainty represents the measured night-to-night variations.
%
% page 5: k_{aer}(lambda) = k_0 * lambda^{-alpha}    (lambda in mum)
% page 6: k_0 = 0.013 +- 0.002, alpha = -1.38 +- 0.06
% --->    k(589 nm) = 0.027
% --->   dk(589 nm) = dk_0 * lambda^{-alpha}   ++  k_0 * lambda^{-alpha} * ln(lambda) * dalpha 
%                   = 0.002 * 2.08    ++    0.013 * 2.08 * 0.53 * 0.06 
%                   = 0.0043
% systematic error from variability at 600 nm, figure 5 and page 4 left last row.
%
%give a mean value of about AOD=0.1 at 550~nm. We further assume a typical
%\AA ngstr\"om index of 1, hence AOD$_{589\,\mathrm{nm}} \approx 0.093$. This value
%can be considered fairly conservative, compared for instance with mean AOD values of 0.04, measured at 
%the Pierre Auger Observatory~\cite{Louedec}, and values AOD$>0.09$ occuring only 10\% of the time. 
%As we will later see, this number enters only in the worst case scenario and is hence justified to be 
%probably a bit higher than during average weather conditions.
%
% 0.027 - 0.005 = 0.022  --> AOD = 0.02
% 
Since the AOD is an integral value over the entire troposphere, we
further need a model for the height distribution of aerosols. First we consider that stratospheric aerosols 
contribute to $k_\mathrm{strato}=0.005$ to the measured aerosol extinction. 
Moreover, we can expect that at
astronomical sites during the night only a
\textit{nocturnal boundary layer} of aerosols extends to about 2000~m
above the ground, possibly showing residuals of the 
day-time \textit{planetary boundary layer} at its edges. A fair approximation
may be dividing the aerosol extinction by 2000~m to derive a \textit{constant aerosol extinction coefficient} 
for heights from ground to 2000~m above it:
\begin{equation}
\beta_\mathrm{aer}(589~\mathrm{nm})_\mathrm{tot} \approx (1.0 \pm 0.3) \cdot 10^{-5}~\mathrm{m^{-1}}~. 
\end{equation}
The extinction coefficient contains an absorption part and a
scattering part; however, for continental clean environments, we can
assume that more than 95\% is due to scattering (i.e. the
\textit{single scattering albedo} is greater than 0.95). Furthermore,
we can use the \textit{Henyey-Greenstein} formula~\cite{Henyey1941}
to estimate the angular distribution of scattered light:
\begin{eqnarray}
\beta_\mathrm{aer}(589~\mathrm{nm},\theta) &\approx& (1.0 \pm 0.3) \cdot 10^{-5} \cdot \frac{1-g^2}{4\pi} \cdot \nonumber\\
 && \bigg(\frac{1}{(1+g^2-2g\cos\theta)^{3/2}} + \nonumber\\
 &&   {} + f \frac{3 \cos^2\theta -1}{2\cdot (1+g^2)^{3/2}} \bigg) ~\mathrm{m^{-1}}~,
\end{eqnarray}
where $g$ varies between 0 and 1 and represents the mean value of
$\cos(\theta)$. The parameter $f$ characterizes the strength of the second component to the backward scattering peak. 
Typical values for clear atmospheres and desert environments are $g \approx 0.6, f \approx 0.4$~\cite{Louedec}. 
%The value is also obtained at the Pierre~Auger Observatory~\cite{Louedec}. 
Inserting these numbers, we obtain a very approximate expression for the aerosol volume scattering 
cross section, valid if the laser light is observed at heights below $\sim$2~km above ground: 
% 
% (1-0.6^2)* (  1/(1+0.6^2-1.2*cos(theta))^(3/2) + 0.4*(3cos(theta)-1)/2*(1+0.6^2)*1.5) = 0.64 * ( 1 /(1.36-1.2*cos(theta))^(3/2) + 0.13*(3*cos(theta)-1) )
%   ~ 0.43*  (  1/(1-cos(theta))^(3/2)  + 0.19*(3*cos(theta)-1) )
% 4.8E-5/4pi*0.95*0.43 = 1.55E-6
% 1.0E-5/4pi*0.95*0.43 = 3.25E-7
%
\begin{eqnarray}
\beta_\mathrm{aer}(589~\mathrm{nm},\theta) &\approx& (3.3 \pm 1.0)\cdot 10^{-7} \cdot \nonumber\\
     &&  \cdot \bigg( \frac{1}{(1-\cos\theta)^{3/2}} + \nonumber\\
     &&  {} + 0.19\cdot (3\cos\theta-1)\bigg)~\mathrm{m^{-1}}~.
\end{eqnarray}
Contrary to the Rayleigh scattering case on molecules, this value can
show large variations, depending on atmospheric conditions. For
instance, a layer of haze can dramatically increase the aerosol
scattering cross section, while scattering at heights above 2~km will
probably result in a negligible contribution. 

\begin{figure}[h!t]
\centering
\vspace{-0.2cm}
\includegraphics[width=0.4\textwidth]{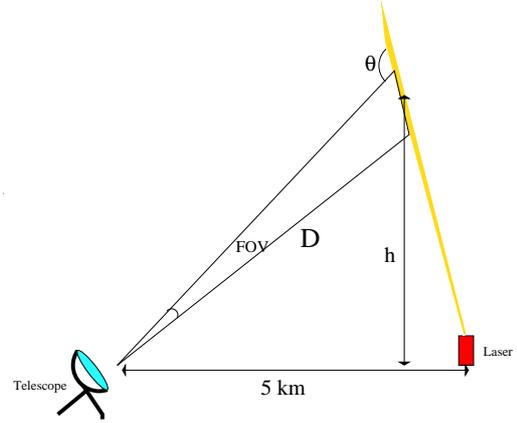} 
\vspace{-0.2cm}
\caption{\label{fig} Sketch of the chosen geometry conventions.}
\end{figure}

\vspace{-0.5cm}
\section{Amount of spurious light in individual CTA pixels}

We can now derive the amount of light observed by a single pixel,
assuming that the camera of a CTA telescope observes the laser beam
at an angle $\theta$ with respect to the optical axis of the
telescope, defined such that if both laser and telescope optical axes
are parallel, then $\theta=\pi$, if the axes cross perpendicularly,
then $\theta=\pi/2$ (see figure~\ref{fig}). Moreover, the laser is sufficiently collimated that
the observed beam width is always smaller than the pixel FOV. The pixel will then observe a part of the laser track,
corresponding to its field-of-view (FOV): 
\begin{equation}
d_\mathrm{track} = \frac{\textit{FOV}_\mathrm{pix}\cdot D}{\sin(\theta)}~,
\end{equation}
where $D$ is the distance of the telescope to the laser beam at the
place where the optical axis and the beam cross. The observed photon flux 
inside one laser's track is:
\begin{equation}
N_\mathrm{laser} = \frac{P_\mathrm{laser}\cdot \lambda}{h\,c} \approx 5.9\cdot10^{19}\quad \mathrm{s^{-1}}~.
%\cdot \frac{\textit{FOV}_\mathrm{pix}\cdot D}{\sin(\theta)}~(\mathrm{m}) 
% \quad.
\end{equation}

Assuming that $D$ is large, we can approximate the scattering angle as
constant throughout the crossing of the beam through the FOV of the
pixel. The  observed laser track will scatter into a solid angle
$\Omega = A_\mathrm{tel} / D^2$, where $A_\mathrm{tel}$ is the area of
the telescope mirror. Assuming  a photo-multiplier quantum efficiency (QE)
at 589~nm $\textit{QE}_{589\,\mathrm{nm}}$ and a combined reflection
and photo-electron (phe) collection efficiency of $\xi\sim 0.9$, we obtain: 
%
% 0.5E-6*(cos^2(theta)+1) + 1.55E-6/(1-cos(theta))^(3/2) = 1.55E-6*(  0.3*(cos^2(theta)+1) + 1/(1-cos(theta))^(3/2)  )
% 
% 1.55E-6 * 5.9E19 * 0.9 =  9.6E13
\begin{eqnarray}
N_\mathrm{pixel} &\approx& N_\mathrm{laser} \cdot \xi \cdot \textit{QE}_{589\,\mathrm{nm}} \cdot \frac{A_\mathrm{tel}}{D^2} \cdot \frac{\textit{FOV}_\mathrm{pix}\cdot D}{\sin(\theta)}\nonumber\\
                   && \cdot \big(\beta_\mathrm{mol}(589~\mathrm{nm},\theta,h) +\beta_\mathrm{aer}(589~\mathrm{nm},\theta,h) \big) \nonumber\\
                &\approx& 8.2\cdot 10^{13} \cdot \textit{QE}_{589\,\mathrm{nm}}  \cdot \frac{\textit{FOV}_\mathrm{pix}\cdot A_\mathrm{tel}}{D\cdot \sin(\theta)} \cdot \nonumber\\
                &&  \cdot \bigg( \big(0.32 - 7.4\cdot 10^{-6} \cdot H\big)^{4.256} \cdot ( \cos^2\theta + 1 )  + \nonumber\\
                &&  + (1 - \Theta(h,2~\mathrm{km})) \cdot \big( \frac{0.21}{(1-\cos\theta)^{3/2}} + \nonumber\\ 
                &&  {} +  0.04\cdot(3\cos\theta-1)   \big) \bigg)~\mathrm{s^{-1}}~,
\label{eq.final}
\end{eqnarray}
where $\Theta(x)$ is the Heaviside function, $H$ is the altitude of the scattering point, i.e. 3~km, and $h$ the height above ground at which the laser is observed.\\

\noindent
Eq.~\ref{eq.final} allows us to draw the following preliminary conclusions:
\begin{enumerate}
\item Those cameras will be affected most which have the highest
  combination of $\textit{FOV}_\mathrm{pix}\cdot A_\mathrm{tel} \cdot \textit{QE}_{589\,\mathrm{nm}}$. 
\item Although the function $(1-\cos^2(\theta))/\sin(\theta)$ has a
  (divergent) maximum at $\theta=0$, i.e. parallel beams, this does not
  mean that a laser beam parallel to the optical telescope axis will
  yield the highest (back)scatter return. Since we have neglected the
  reduction of the solid angle with increasing distance, this is an
  artificial effect of Eq.~\ref{eq.final}. However, at $\theta =
  0.9\pi$, $(1-\cos^2(\theta))/\sin(\theta)$ has only increased by a
  factor of two, suggesting that Eq.~\ref{eq.final} is at least not valid
  for viewing angles higher than that value. 
\item Apart from the effect described in point 2, the distance $D$ to
  the laser beam reduces the amount of registered light linearly.
  This is due to the combination of reduced solid angle (which
  goes with $D^{-2}$) and the increased part of the track spanned by
  the FOV of a pixel (which goes with $D$, due to the one-dimensional
  propagation of the laser beam). 
\end{enumerate}

\vspace{-0.5cm}
\section{Case scenarios}
The following table gives an overview of the relevant parameters of
each CTA camera under consideration: 
\begin{center}
\begin{tabular}{lccc}
Camera  & Mirror area & Pixel FOV & $A \cdot$FOV \\
        &  (m$^2$)   &   (mrad)  & $(\mathrm{m}^2 \cdot\mathrm{rad})$\\    
\hline
LST      &  415       & 1.7   &  0.71 \\
MST-DC   &  113       & 3     &  0.34 \\
MST-SC   &  $\sim$50  & 1.17  &  0.06 \\
SST-DC   &  12.5      & 4.14  &  0.05 \\
SST-SC   &  $\sim$6   & 2.9   &  0.02 \\
\hline
\end{tabular}
\end{center}
%
% To calculate the table value, I used the following values
% - for LST pixel of 5 cm and F=28m
% - for MST-DC pixel of 5 cm and F=m
% - for MST-SC values provided by Vladimir
% - for SST-SC values provided by Canestrari
% - for SST-DC I considered the Winston cone d=2.32cm and F=5.6m
%
One can see that the highest spurious signal is expected for the
LSTs, except the distance from one MST to the laser beam is a factor
two smaller than the one from the LSTs. The E-ELT lasers are
located at a distance of 5~km from the CTA, but the largest distance that
an MST can have w.r.t. the LSTs, is only 1~km, the worst-case distance of an MST to E-ELT would be 4~km. 
Since the effect of a decrease in distance by a factor 0.8 is much less than an increase of $A \cdot \textit{FOV}$ by a factor 2.1, 
we consider only the LST case further.  

%\subsection*{Worst case scenario}
\vspace{3mm}
\noindent
{\bf Worst case scenario}
We remark that this scenario is highly improbable to happen. 
The laser is shooting at 70$^\circ$ zenith angle towards the CTA, the
LSTs are looking into the direction of the laser, at 20$^\circ$
zenith angle.  The distance to the laser beam is then as low as
1700~m, scattering occurs at 4500~m a.s.l., aerosol scattering takes
place and increases the amount of scattered light. 
In this case, the scattering angle $\theta$ is 90 degrees and: 
%
% old:  8.2E13*0.71 / 1700 * (0.32*(1-2.3E-5*4500)^4.256 +  1-0.19) =
% new: 8.2E13*0.71 / 1700 * (0.32*(1-2.3E-5*4500)^4.256 + 0.21-0.04) = 
% new2: 8.2E13*0.71 / 1700 * (0.32*(1-2.3E-5*4500)^4.256 + 0.25 +- 0.1) = 
% 
\begin{equation}
N_\mathrm{pixel} = (13 \pm 3) \cdot \textit{QE}_{589\,\mathrm{nm}} \cdot N_\mathrm{lasers}~~\mathrm{ph.e. / ns}~,
\end{equation}

However, in this case, the beam will likely appear broader in the camera, since seen out of focus. Therefore, 
the amount of light will appear rather distributed among a row of two to three pixels.

\vspace{3mm}
\noindent
{\bf A typical bad case} 
Both ELT and the CTA observe at 45$^\circ$ zenith angle, with the laser
pointing towards the CTA. The distance to the laser beam is then 3500~m,
scattering occurs at 5500~m~a.s.l., aerosol scattering can be
disregaded. In this case, the scattering angle $\theta$ is again 90
degrees and:  
\begin{equation}
N_\mathrm{pixel} = 3.3 \cdot \textit{QE}_{589\,\mathrm{nm}}\cdot N_\mathrm{lasers}~~\mathrm{ph.e. / ns}~,
\end{equation}

\vspace{3mm}
\noindent
{\bf A typical good case. } 
The CTA observes at 30$^\circ$ zenith angle, the lasers pointing upwards.  
The distance to the laser beam is then 10000~m, scattering occurs at
11700~m a.s.l., aerosol scattering can be disregarded. 
In this case, the scattering angle $\theta$ is 120 degrees and: 
\begin{equation}
N_\mathrm{pixel} = 0.8 \cdot \textit{QE}_{589\,\mathrm{nm}}\cdot N_\mathrm{lasers}~~\mathrm{ph.e. / ns}~,
\end{equation}

\begin{figure}
\centering
\includegraphics[width=0.45\textwidth]{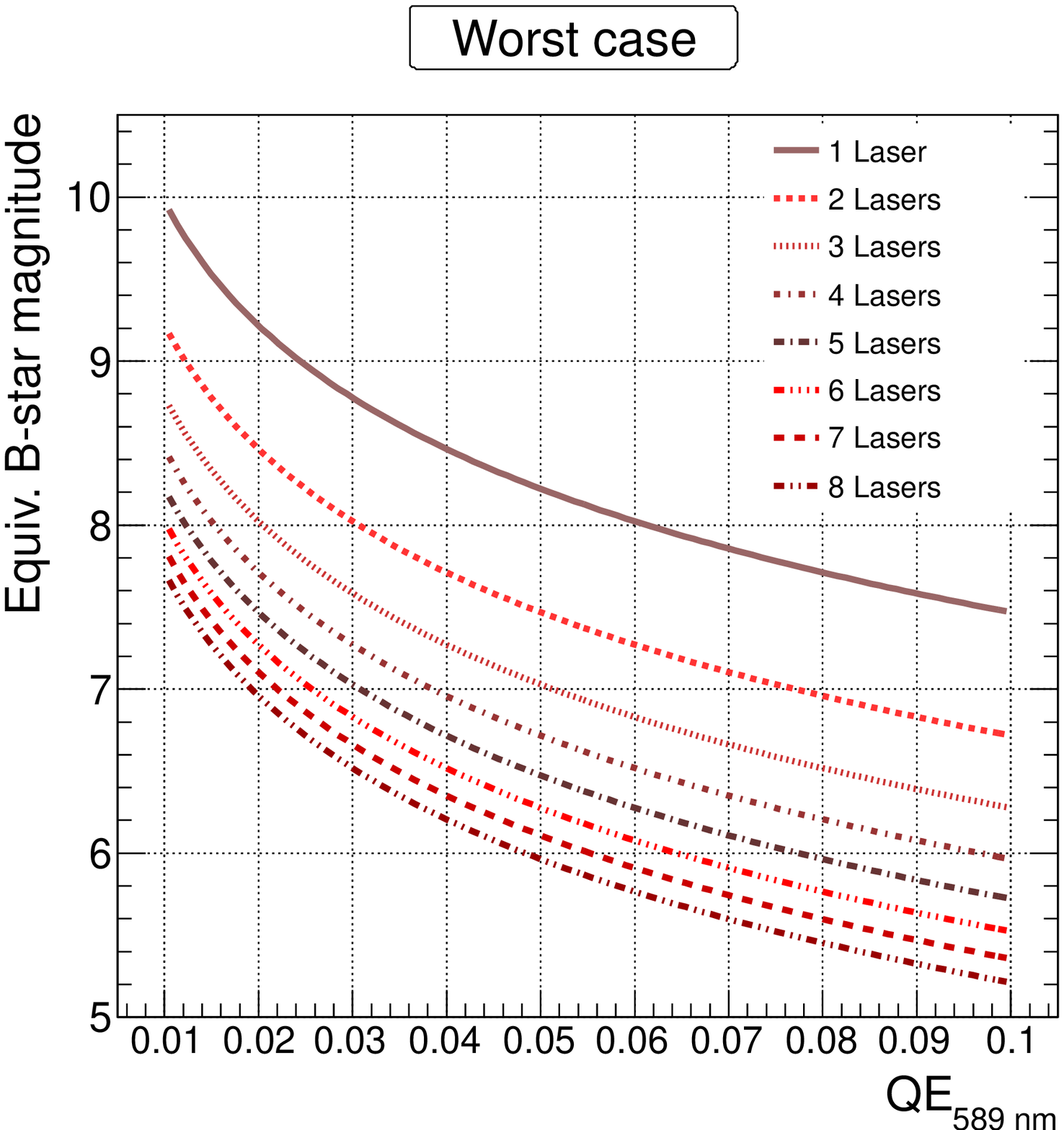}
\includegraphics[width=0.45\textwidth]{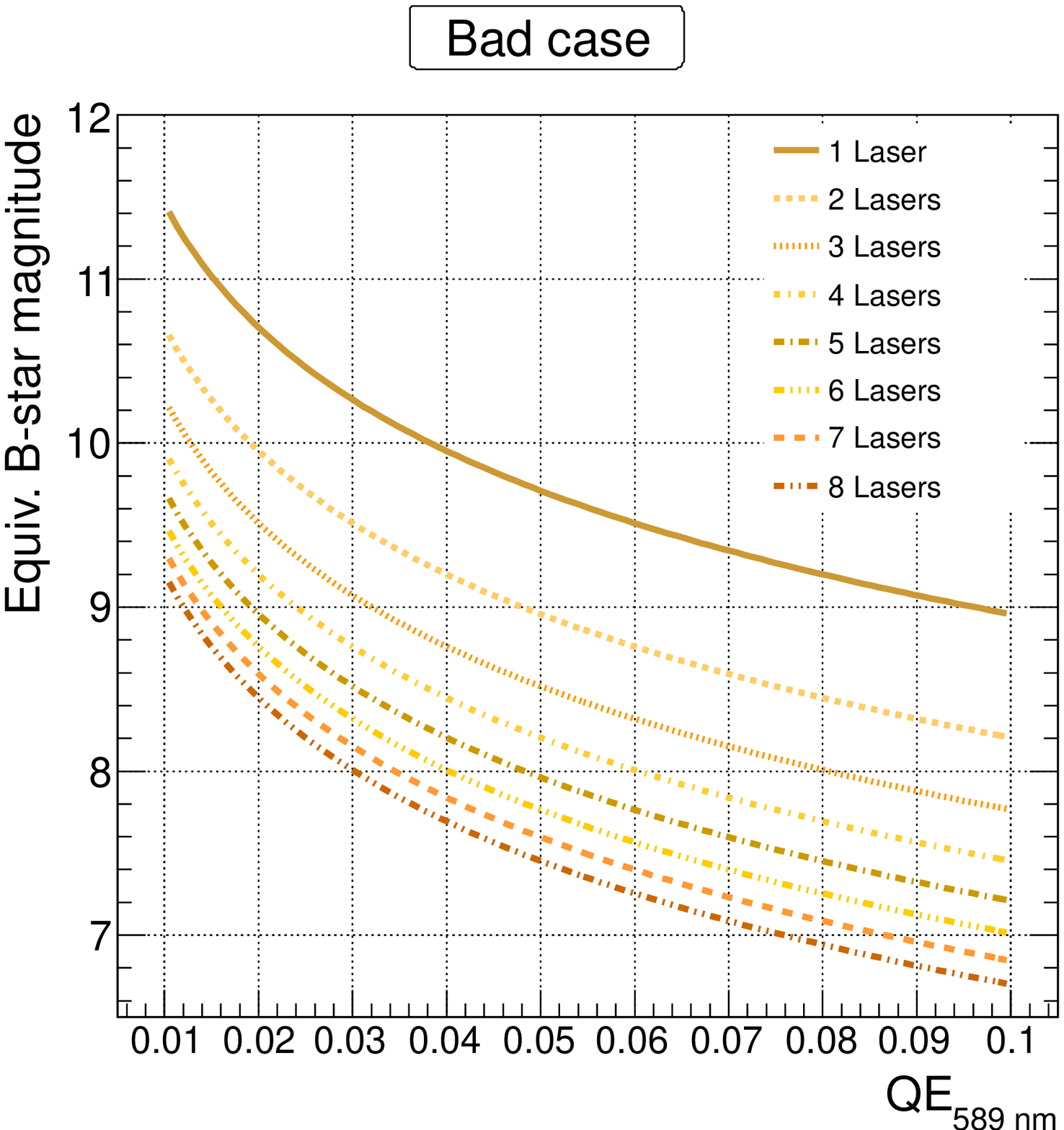}
\includegraphics[width=0.45\textwidth]{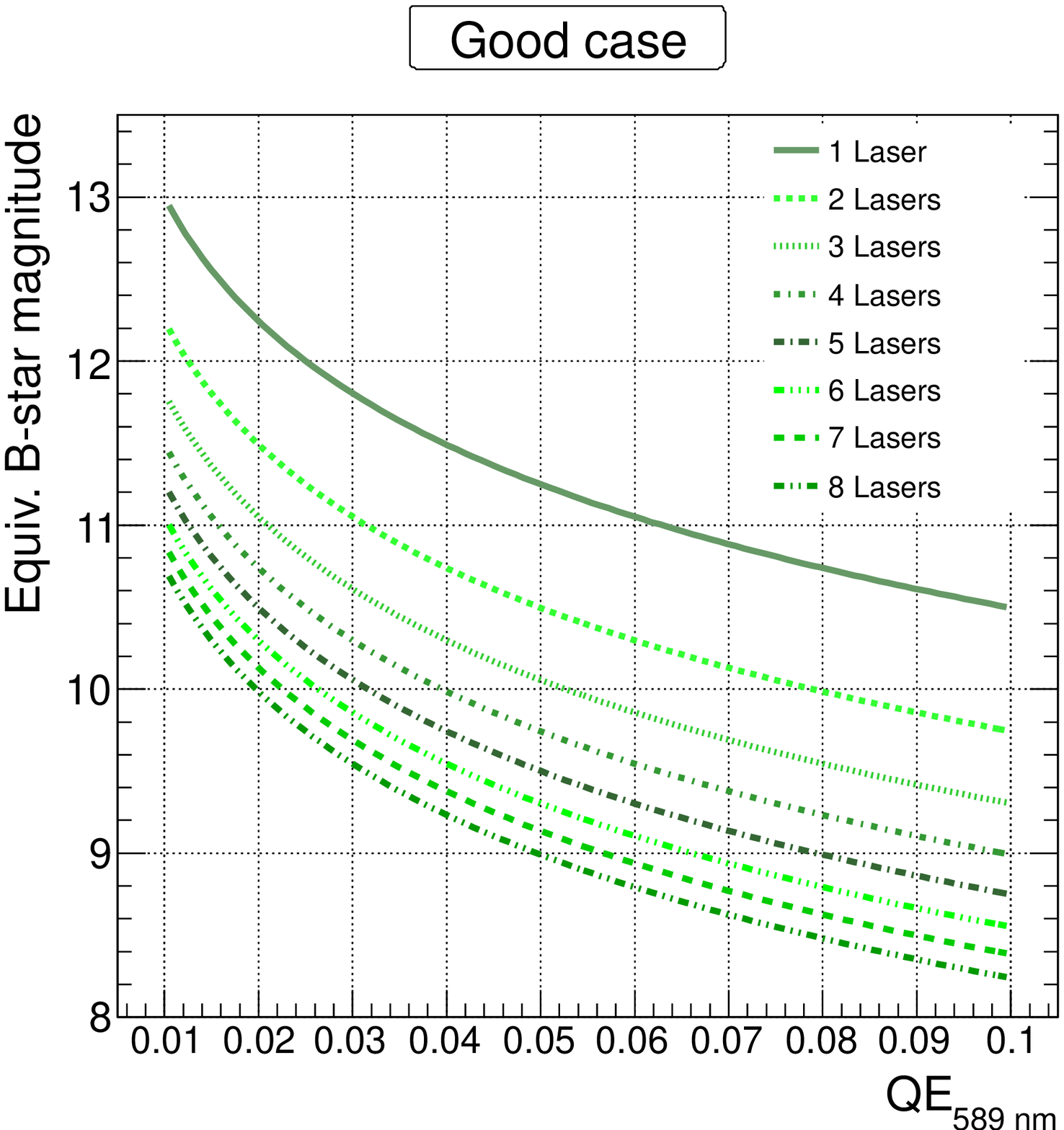}
\caption{\label{fig2}Equivalent star
  $B$-magnitude for LST pixels for different QE values at
  589~nm, and for three different pointing scenarios. The different lines represent the number of lasers 
  observed at the same time in one camera pixel.}
\end{figure}

\noindent
Figure~\ref{fig2} shows the equivalent $B$ star magnitude in one pixel vs. the photomultiplier QE at the laser wavelength. 
The magnitudes have been derived assuming the flux of Vega from~\cite{Bohlin} and a common photomultiplier 
$\mathrm{QE}_{440\,\mathrm{nm}}=0.35$ at 440~nm wavelength, and a spectral width of the QE of $d\lambda/\lambda=0.2$. 
A global atmospheric extinction of $z=0.25$ is further assumed for the $B$-filter.

\vspace{-0.3cm}
\section{Discussion and Conclusions}

%% LONS: MAstro::LONSFluxJy(MAstro::kAtmoscopeB)*0.3*1.5E7*1E-9*415*pow(1.7E-3*TMath::RadToDeg()/2.,2)*TMath::Pi()
% 

We expect from negotiations with the provider that the QEs of the photomultipliers, used for the LST, will
oscillate between roughly 2\% and 7\% at 590~nm wavelength. 
While the night sky background at the Chilean site is expected to produce
roughly 0.2~phe/ns per LST pixel, 
% 0.8*0.02 = 0.016
% 0.8*0.07 = 0.06
the expected signal seen from an E-ELT lasers for a typical good case is 0.02~phe/ns in the case of photo-multipliers with less 
than 2\% QE at 589~nm and 0.06~phe/ns for the case of 7\%~QE.
In this situation, the laser photons should not affect data significantly, since their amount is much lower than the light-of-night sky.
%unless the trigger thresholds are adapted for those pixels
% which view the laser beam. 
The typical bad case, together with 2\% QE, corresponds to the typical good case with 7\% QE and should 
still be acceptable, both in terms of spurious islands in the image and in terms of
fake triggers. This possibility is rather small, since the ultra-collimated 
laser light is seen far from the telesopes, and should hit pixels in one single row. Therefore any next-neighbor
logic of the trigger, if applied,  should reject these events. 

However, having rows of 8$^m$ stars throughout the camera, as would typical for the bad case with 7\% QE,   
could compromise the analysis already and should be avoided. 
A completely different case is the absolute worst case situation:  
here, we can get more than an order of magnitude higher phe-fluxes than the night-sky background, and the images 
would be strongly affected.  
Both cases should be avoided, but should not occur frequently, since the E-ELT lasers must point to low elevations, and towards the CTA for this to occur. 

Another possible solution would be to exclude the pixels in the
camera hit by the laser light in the reconstruction
software, given that the laser direction should be known at any moment. Since the laser light is received during each shower event, 
the track can affect the direction and energy reconstruction of the event.

%\vspace{-0.3cm}
%\section{Conclusion}

If the numbers plugged into formula~\ref{eq.final} are correct, i.e. if the core of CTA lies at least 5~km from the site of the \mbox{E-ELT}, and 
if LST QEs are below 7\% at 589~nm (better even $<2$\%), and not more than 8 lasers are fired synchronously at the E-ELT, 
then the effect of the laser tracks in the CTA cameras should be sufficiently small in typical situations that it will not compromise trigger rates 
and image reconstruction. 

Only the bad and worst case situation will have an effect on the recorded images and should be avoided, especially if the photo-multiplier QE is 
higher than 2\% at 589~nm. The worst case (E-ELT lasers pointing at very low elevation towards CTA), must be avoided. 
However, this should happen so rarely that it can probably be avoided by coordinated scheduling.  

%With this information in mind, we show that the E-ELT lasers do not affect the selection of Chile as potential candidate site for CTA-South. 

\vspace{0.5cm}
{\bf Acknowledgements:\xspace} 
We gratefully acknowledge support from the agencies and organizations  
listed in this page: \url{http://www.cta-observatory.org/?q=node/22}.
\noindent

\bibliographystyle{plain}

\end{document}